\documentclass[final,5p,11pt]{elsarticle}

\usepackage{graphicx}
\usepackage{amssymb, amsmath,bm}

\usepackage{lineno}
\usepackage{xcolor}

\journal{Acta Materialia}

\begin{document}

\begin{frontmatter}

\title{Scratching the surface: Elastic rotations beneath nanoscratch and nanoindentation tests}



\author[materials]{A.Kareer}
\author[eng,materials]{E. Tarleton}
\author[Culham,materials]{C. Hardie}
\author[Aston]{S.V.Hainsworth} 
\author[materials]{A.J.Wilkinson}

\address[materials]{Department of Materials, University of Oxford, Parks Road, Oxford, OX1 3PH, United Kingdom}
\address[eng]{Department of Engineering Science, University of Oxford, Parks Road, Oxford, OX1 3PJ, United Kingdom}
\address[Culham]{Culham Centre for Fusion Energy, UK Atomic Energy Authority, Abingdon,
OX14 3DB, United Kingdom}
\address[Aston]{Aston Institute of Materials Research, School of Engineering and Applied Science, Aston University, Aston, Birmingham, B4 7ET, United Kingdom}

\begin{abstract}

In this paper, we investigate the residual deformation field in the vicinity of nanoscratch tests using two orientations of a Berkovich tip on an (001) Cu single crystal. We compare the deformation with that from indentation, in an attempt to understand the mechanisms of deformation in tangential sliding. The lattice rotation fields are mapped experimentally using high-resolution electron backscatter diffraction (HR-EBSD) on cross-sections prepared using focused ion beam (FIB). A physically-based crystal plasticity finite element model (CPFEM) is used to simulate the lattice rotation fields, and provide insight into the 3D rotation field surrounding a nano-scratch experiment, as it transitions from an initial static indentation to a steady-state scratch. The CPFEM simulations capture the experimental rotation fields with good fidelity, and show how the rotations about the scratch direction are reversed as the indenter moves away from the initial indentation.
\end{abstract}


\end{frontmatter}


\section{Introduction}

The extensive development of nanomechanical testing instruments has expanded the capabilities of nano-scale measurement beyond basic indentation hardness. Nanoscratch has generated significant interest and can be performed on commercial nanoindenters, requiring only minor adaptations to the software. When used in combination, nanoscratch and nanoindentation provide a powerful means to investigate the near surface mechanical and tribological properties of small volumes of material \cite{xia2006nanoindentation, tsui1995nanoindentation,beake2002nanoindentation, Beake2013}; it enables the study of nanoscale friction \cite{lafaye2006friction, bhushan1995friction} and allows the adhesive strength and fracture properties of coated systems to be characterised \cite{hainsworth_bull_page_1998, zhang2005nanoindentation, CHANG20103307}. 

Macroscopically, Tabor showed that on the basis of plastic deformation, there is a strong correlation between the indentation and scratch hardness when the measurement is based on a mean pressure \cite{Tabor_1956}. As a consequence, the indentation hardness is the key metric used to define wear resistance \cite{Archard1986}. The complex surface interactions involved in wear processes are extremely difficult to understand and simplified models such as this do not completely take into account the physical mechanics that are occurring. Nanoscratch testing has the advantage, in that it provides an experimental platform to reproduce a single point, sliding asperity contact, that is believed to control the process of abrasive wear \cite{Beake2013, Bakshi2011, Hutchings1992}. Hence, it is becoming increasingly common to use this technique to study the near surface mechanical properties via determination of the scratch hardness, from which the wear resistance of material systems can be inferred \cite{Graça2008, Brookes_1972, BULL200699, SCHUH2002735}. The remaining issue resides in the definition of scratch hardness; the simplest and most frequently used measurement of scratch hardness is analogous to indentation hardness and is defined as the ratio between the normal load and the projected load bearing area. The studies that have used this definition, show substantial differences between the measured indentation and scratch hardness \cite{MAAN1977365, Useinov2012, Williams1996}. By further incorporating the lateral force into the measurement, it is possible to obtain a scratch hardness measurement that is in closer agreement to the indentation hardness, on isotropic materials \cite{tayebi_conry_polycarpou_2003, tayebi_conry_polycarpou_2004}. This however, is not applicable to all material classes, particularly metallic samples that exhibit work hardening and anisotropy, as shown in \cite{Kareer2016, Kareer2016paper2}. A number of critical considerations must additionally be accounted for; namely the effect of friction, plasticity size effects, the resultant strain on the material, work hardening as a result of evolving dislocation structure and the direction of flow of displaced material in each loading direction \cite{Brookes_1972}. In order to interpret the differences measured in indentation and scratch hardness, it is important to develop a deeper understanding of the mechanics of nanoscratch formation.  
 
Experimental observations of the deformation field beneath indentation experiments, have been used to interpret the hardening behaviour in various materials. These studies have revealed that the plastic zone is extremely complex and there is a continued effort to relate the deformation field to the measured mechanical properties \cite{Zhang2014, Kiener2006, Kysar2007, Gao2015, Demir2009}. Several studies have reported that for indentation with a geometrically self-similar indenter, the geometrically necessary dislocation (GND) structure does not develop in a self-similar, hemispherical way as often assumed in simplified explanations \cite{Rester2009, Rester2007, Wilkinson2010}. The lattice rotation fields below indentation experiments with various tip geometries, have revealed distinct patterning within the plastic zone that exhibit well defined boundaries and a steep orientation gradient where a change in the sign of the rotation direction is observed \cite{Zhang2014, Demiral2014, Zaafarani2006}. The investigation of plastic deformation and induced lattice rotations is of great interest for an improved micromechanical understanding of indentation experiments owing to the close connection between crystallographic shear and the resulting lattice rotation.

Simulation methods, such as the Crystal Plasticity Finite Element Method (CPFEM), provide further insight into the mechanics of the deformation field, when applied in conjunction with indentation experiments \cite{Dunne2012}. Through incorporation of an appropriate, physically based, constitutive model along with details of the microstructure and constitutive parameters, CPFEM enables the effect of grain size, crystallographic orientation \cite{Kucharski2014} and plasticity size effects \cite{Liu2015} to be studied. Zaafarani and co-workers simulated the lattice rotation field below spherical indentations in copper using CPFEM, and directly compared the simulated rotation fields with that obtained from EBSD \cite{Zaafarani2006}. Simulating the lattice rotation field facilitates the interpretation of the deformation mechanisms, by separating the crystallographic shear occurring on individual slip systems, and directly relating it to the patterns observed in the lattice rotation fields \cite{Zaafarani2008}. CPFEM has the added benefit in that it offers information on the spatial 3D distribution of the deformation field, in real time, as it evolves throughout the experiment. This is necessary to interpret dynamic experiments, whereby experimentally it is only possible to study the final, deformed state via postmortem analysis.
 
 In comparison to nanoindentation experiments, the study of deformation below nanoscratch experiments is still in its infancy; the deformation field is further complicated due to the lateral force. Macroscopic scratch experiments show that plastic deformation induces changes in the microstructure of the material, resulting in a distinct discontinuity between a surface layer and the underlying bulk material \cite{Greiner2016, Sundaram2012, Romero2014, Li2012, Emge2009, Bednar1995}. This physical boundary has also been identified in TEM studies around nanoscratch experiments in Ni$_3$Al where the plastic zone consists of a core region with high dislocation density surrounded by an outer region with lower dislocation density \cite{Wo2008}.
 
 Simplified mechanistic models have been proposed to simulate the plasticity dominated deformation field around nanoscratch experiments similar to those used for nanoindentation that assume the plastic zone is proportional to the scratch width \cite{Williams1996, Burnett1988} but these models are purely theoretical and, in most cases, are not validated. Isotropic Finite Element models have been used to simulate the strain field around nanoscratch experiments in an attempt to validate analytical models \cite{lee2008relationship, bellemare2008effects} and to describe the strain field in both bulk and coated systems \cite{bucaille2002finite,perne2014experimental}. A model by Holmberg and co-workers found that the stress field under scratch experiments in coated systems is different to that under bulk samples. This was attributed to the mismatch between material properties in the coating and substrate, which restricted the ability for the residual stresses to elastically recover. In the absence of a coating, elastic recovery is accommodated, resulting in a different stress field \cite{holmberg2006tribological}. However, a detailed study of the effect of crystallographic orientation and the resulting lattice rotation field surrounding nanoscratch experiments remains unexplored. 

In this paper, we use High Resolution EBSD (HR-EBSD) to experimentally map the lattice rotation field in the vicinity of nanoindentation and nanoscratch experiments \cite{Wilkinson2010, Britton2017, Britton2013, Wilkinson2014, Jiang2015}. Scratch and indentations were generated under the same normal force (3mN) using a Berkovich indenter, in single crystal copper. A physically based CPFE model is used to simulate the scratch experiment. Lattice rotation fields from the simulation are directly compared with the experimental results and provide an insight into the three-dimensional mechanisms that occur during deformation beneath a sliding contact which can help understand the quantitative differences observed between indentation and scratch hardness. 

\section{Methods}
\subsection{Nanoscratch and nanoindentation }

\begin{figure*}[htb]
\center
\includegraphics[width=1\linewidth]{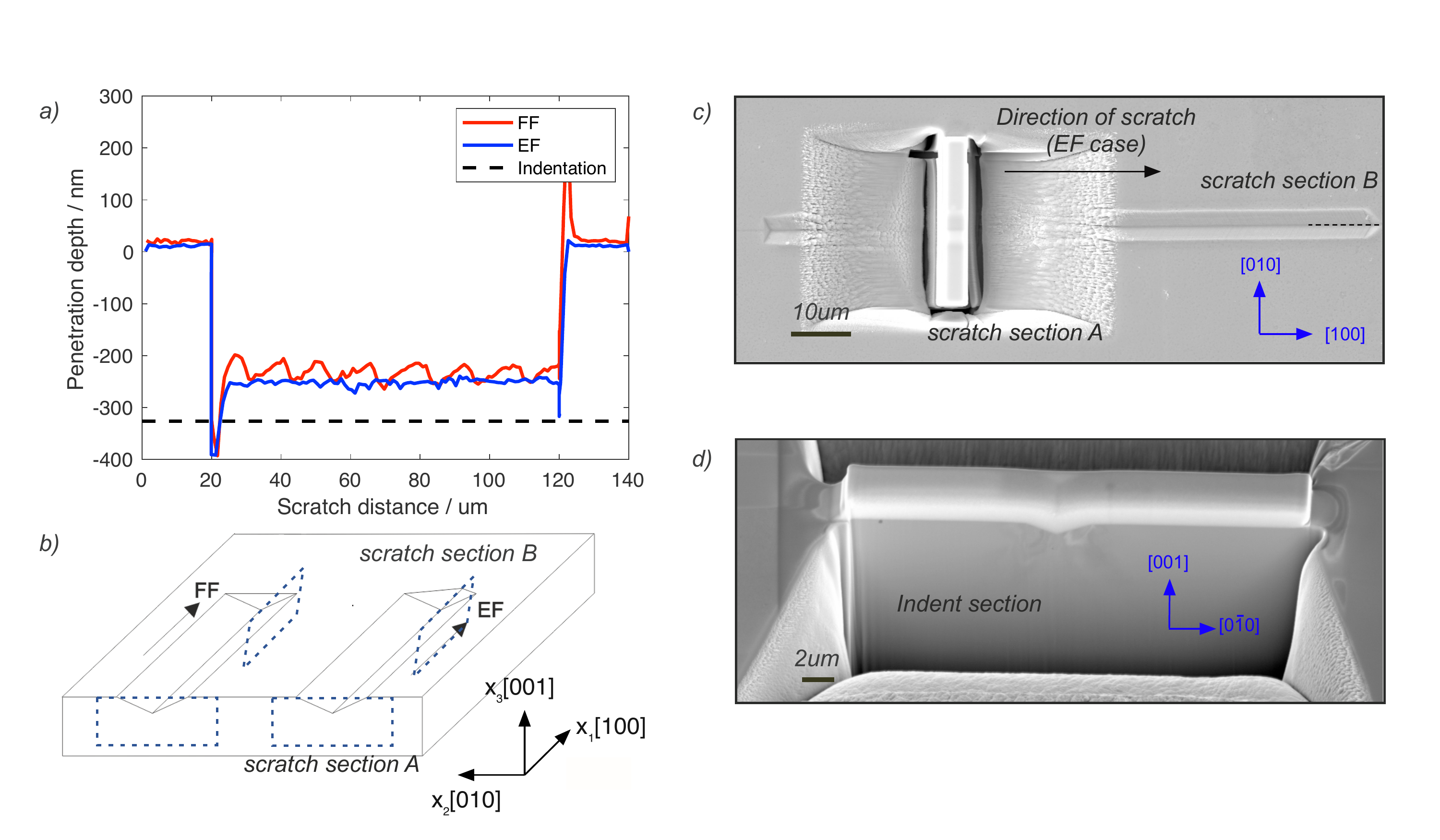} 
\caption{a) Raw experimental penetration depth vs. scratch distance for FF and EF scratch. b) Schematic of the scratch directions with respect to the crystallographic orientation and location of sections lifted for EBSD analysis. c) SEM image of EF scratch and FIB prepared scratch section A d) SEM image of FIB prepared indentation cross-section.}
\label{Figure1}
\end{figure*}

Nanoscratch and nanoindentation experiments were carried out on a sample of single-crystal, oxygen-free pure copper with 99.9\% purity, oriented in the (001) crystallographic plane (obtained from Goodfellow UK). The sample was annealed in air for 4 hours at $600^\circ$C, followed by a mechanical and electrolytic polish in order to obtain a smooth flat surface, with negligible residual stresses. The indentation and scratches were made using a Keysight (formerly Agilent, formerly MTS) G200 instrumented indentation system, fitted with a lateral force measurement probe and a Berkovich diamond tip. {\color{black} The tip radius of 20nm was measured using AFM. The indentation was made using a quasi-static loading function to a maximum normal force of 3 mN and the indenter was aligned such that one facet of the indenter was perpendicular to the [100] direction}. The scratches were formed in both the edge forward (EF) and face forward (FF) tip orientation, parallel to the [100] direction (corresponding to the x$_1$ direction in Figure \ref{Figure1}(b)) at a velocity of 10 $\mu$m s$^{-1}$. {\color{black}A scratch length of 100 $\mu$m} and constant normal force of 3 mN, were used for the scratches and a ‘three-pass’ scratch method was implemented to correct for surface roughness and sample tilt, {\color{black}by subtracting the surface profile displacement (pass 1) from the scratch displacement (pass 2). The final profiling pass (pass 3), offers insights into elastic recovery. Full details of the three-pass scratch method are provided in \cite{Kareer2016paper2}.} The corrected penetration depth channel is plotted as a function of scratch distance in Figure \ref{Figure1}(a). 

\subsection{Cross-sectioning}
Cross-sections through the nanoindent and the scratches were prepared using a Zeiss Auriga FIB-SEM. Cross-sections were taken from two locations in the scratch; scratch section A is cut across the centre of the scratch and scratch section B is cut at the end of the scratch (these locations are shown in Figure \ref{Figure1}(b)). The indentation section was taken from the centre of the indentation, in the same orientation as the scratch such that the cross-sectioned surface was oriented in the (100) crystallographic plane. Cross-sectional slices, of approximately $20\mu \textrm{m} \times 10\mu \textrm{m} \times 3\mu\textrm{m}$ in size, were lifted from the sample in-situ and mounted on an Omniprobe TEM copper grid. SEM images of the prepared cross-sections, prior to lift-out, are given in Figures \ref{Figure1}(c) and \ref{Figure1}(d). 

\subsection{High Resolution Electron Backscatter Diffraction}
EBSD measurements of the cross-sections and surface of the scratches were made in a Zeiss Merlin FEG SEM equipped with a Bruker e$^{-}$Flash$^\textrm{HR}$ EBSD detector operated by Esprit 2.0 software. The EBSD patterns (EBSPs) were acquired using an electron beam energy of 20kV and a probe current of 5nA; a step size of 50nm was used and EBSPs were collected and saved at a resolution of $800 \times 600$ pixels. 

HR-EBSD is used to map the lattice rotation field in the vicinity of the indentation and scratch experiments. The technique uses an image cross-correlation based analysis to measure the lattice curvature within the crystal. A reference pattern is selected from a location within the grain far from the regions of high deformation (in the case of the single crystal used in this study, the reference pattern is selected far from the indented/scratched region of the orientation map). Each test pattern within the map is cross-correlated with respect to the reference pattern; the pattern shifts of a number of regions of interest (ROIs) are measured and related to the crystal lattice rotation. A full description of the mathematics describing the method can be found in references \cite{wilkinson2006high,Wilkinson2010, britton2012high}. Lattice rotations between test and reference pattern can be used to estimate the GND density based on Nye's framework \cite{nye1953some, Britton2017}. The total dislocation density comprises individual dislocation densities from each dislocation slip system. For an FCC crystal, it is assumed that the GNDs are either pure screw dislocations, or pure edge dislocations with $\langle110\rangle$ Burgers vectors giving 18 types of unknown dislocation densities. 2D maps allow the lattice rotation gradients along two orthogonal axes within the surface to be determined giving six out of the nine lattice rotation tensor components. At each point in the map, a set of possible GND combinations that satisfy the six measured lattice curvatures are found and the combination that gives the minimum total line energy is chosen. Further information on the calculation of GND density can be found in \cite{Wilkinson2010}. For the experimental conditions used, this method measures lattice rotations with an angular sensitivity of $3\times 10^{-4}$ rad which corresponds to a lower bound GND density noise floor of $1.5\times 10^{13} m^{-2}$.

\subsection{Crystal plasticity finite element modelling }
Finite element simulations were performed using Abaqus 2016 to investigate the influence of deformation due to scratching with the EF tip geometry, in single crystal copper. A crystal plasticity user material (UMAT) for Abaqus was used based on the user element (UEL) by Dunne et al. \cite{Dunne2007, Dunne2012}. 

The deformation is decomposed multiplicatively into a plastic, $\bm{F}^p$, and elastic, $\bm{F}^e$, deformation gradient
\begin{equation}
\label{eq:F}
F_{ij} = F_{ik}^e F_{kj}^p 
\end{equation}
the flow rule has the form  
\begin{equation}
\dot{F}_{ij}^p=L_{ik}^p F_{kj}^p.
\end{equation}
Where the plastic velocity gradient, $\bm{L}^p$, is given by the crystallographic strain rate resulting from dislocation glide on the active slip systems with slip direction $\boldsymbol{s}^k$ and slip plane normal $\boldsymbol{n}^k$
\begin{equation}
\label{eq:Lp}
L_{ij}^p = \sum_{k=1}^{k=12} \dot{\gamma}^{k}(\tau) s_{i}^k n_{j}^k
\end{equation}
The crystallographic slip rate $\dot{\gamma}^α$ is given by
\begin{equation}
\label{eq:gammadot}
\dot{\gamma}^{k}(\tau)= A\sinh\left(B(\vert\tau^{k}\vert - \tau_c)\right)\textrm{sgn}(\tau^k)
\end{equation}
for $\vert\tau^{k}\vert > \tau_c$ and $\dot{\gamma}^{k}=0$ otherwise, where the resolved shear stress on slip system $k$ is $\tau^k =\sigma_{ij}n_{i}^k s_{j}^k$. The critically resolved shear stress is assumed the same for each slip system,
\begin{equation}
\label{eq:tauc}
\tau_c(\rho)= \tau_c^0 + C G b \sqrt{\rho}.
\end{equation}
For simplicity we assume that the dislocation density, $\rho$ is proportional to the plastic strain 
\begin{equation}
\label{eq:rhossd}
\dot{\rho} = D \sqrt{\frac{2}{3}\dot{\varepsilon}_{ij}^p\dot{\varepsilon}_{ij}^p}
\end{equation}
where the plastic strain rate $\dot{\varepsilon}^p_{ij}$ is the symmetric part of $L^p_{ij}$. The fitting constants were $A = 10^{-6}$ s$^{-1}$, $B=0.1$ MPa$^{-1}$, an obstacle strength term $C=0.05$ and $D=2.45\times 10^{4}~\mu$m$^{-2}$. Initial values were taken from \cite{cackett2019spherical} and calibrated to match the normal force during the scratch. Consequently plastic deformation induces a dislocation density which hardens the slip systems via an increase in the CRSS, $\tau_c(\rho)$.

{\color{black}Lattice Rotation:
The elastic distortion and rigid body rotation of the lattice is found by computing the elastic part of the deformation gradient, from (\ref{eq:F}):
\begin{equation}
F_{ij}^e={F}_{ik}{F_{kj}^p}^{-1}
\end{equation}
and from this the elastic part of the velocity gradient and its anti-symmetric spin component are calculated:
\begin{align}
L_{ij}^e &=\dot{F}_{ik}^e F_{kj}^{e-1} \\
W_{ij}^e &=\frac{1}{2}(L_{ij}^e - L_{ji}^e)
\end{align}

Finally the crystal orientation matrix $R$ is calculated using an implicit integration of the elastic spin, which is updated at the end of each increment:
\begin{equation}
R_{ij}(t+\Delta t)=\left[\bm{I}-\Delta t \bm{W}^e(t+\Delta t)\right]_{ik}^{-1}R_{kj}(t)
\end{equation}
$R_{ij}$ rotates a vector expressed in the crystal frame to a vector in the deformed material frame. The crystal frame in this study is initially aligned with the reference frame of the model and so $R_{ij}(0)=\delta_{ij}$.

The rotation matrix at the end of the simulation was used to define the elastic rotations about the 3 axes in the reference frame x$_3$, x$_2$ and x$_1$ for comparison with experimental HR-EBSD data: \cite{slabaugh1999computing,das2018consistent}: 
\begin{align}
      \omega_{32} &= \tan^{-1}(R_{32}/R_{33}) \\
      \omega_{13} &= -\sin^{-1}{(R_{31})}  \\
      \omega_{21} &= \tan^{-1}(R_{21}/R_{11}) 
\end{align}
}

A total of 4950 linear hexahedral elements with 8 Gauss points per element (C3D8) were used to represent a block of copper with dimensions of L = 25$\times 15\times 7.5$ $\mu$m, with symmetry boundary conditions applied along the $(010)$ mid plane allowing only half of the domain to be simulated. The scratch test was simulated by modelling an indentation followed by an edge forward (EF) scratch step with a constant displacement of $u_3=-247$ nm and applied lateral displacement of $u_1=10~\mu$m at a rate of 10 $\mu$m/s, these parameters were chosen to match the experiment. {\color{black} Displacement boundary conditions were used to improve the stability and efficiency of the implicit solution for global equilibrium of nodal forces. Force boundary conditions are particularly unstable for highly non-linear problems such as contact.} A scratch length of 10 $\mu$m was found to be sufficient to reach a steady state scratch formation. A biased mesh under the scratch was used for improved accuracy and computational efficiency, with an approximate element size of $w_1 = 0.4~\mu$m along the scratch direction x$_1$, $w_2 = 0.2~\mu$m along x$_2$, increasing up to $w_2 = 1 ~\mu$m far from the scratch. The indenter tip was modelled as a rigid part with a perfect Berkovich geometry. The finite sliding, node to surface, Abaqus contact algorithm was used with the default hard contact property. The absolute values for the lateral and normal force are determined by the material model however, their ratio is governed entirely by the friction behaviour. A friction coefficient of 0.15 was used to specify tangential behaviour between the surfaces in contact. This value was calculated from the experimental data by resolving the normal, $F_{N}$, and tangential forces, $F_{T}$, on the indenter tip faces during sliding; where the friction coefficient is $\mu=F_{T}/F_{R}$. {\color{black}The result is the true friction coefficient between the contact surfaces which is independent of geometry, scratch speed or scratch direction in this case. Therefore values for both tip orientations studied were identical within the experimental scatter. In contrast calculating the friction coefficient without accounting for the geometry, i.e. simply taking the ratio of the applied lateral and normal forces, $\mu=F_L/F_N$, is no longer fully defined by the two materials but is dependent on the stress state and the resultant plastic deformation imposed by the tip orientation/geometry/speed. Consequently $F_L/F_N$ is significantly different for the EF and FF orientations in this study; 0.4 and  0.6 respectively. 

The new friction coefficient described in  \ref{sec:appendix}, $\mu=F_T/F_R$ where $F_T$ and $F_R$ are the tangential and resolved orthogonal  forces acting on the facet(s) of the indenter, depends only on the interaction between the two material surfaces in contact and does not vary with tip orientation. The friction coefficient was found to be 0.15 in both scratch directions and was used as the friction coefficent in the Abaqus surface contact algorithm in the FE model.} The symmetry plane was fixed from translation in the normal direction, x$_2=0$, the nodes on the top surface were traction free, while the remaining four surfaces of the cube were fixed. Elastic anisotropy was used with the following elastic constants for copper: $c_{11}=168.4$, $c_{12}=121.4$, $c_{44}=75.4$ or $E = 66.7$ GPa, $G=75.4$ GPa, and $\nu= 0.419$. 12 $\langle1\bar{1}0\rangle\{111\}$ fcc slip systems were included with an initial CRSS of $\tau_c^0 = 1$ MPa. Further details on the UMAT can be found in \cite{cackett2019spherical,grilli2020characterisation,grilli2020crystal,roberts2020tension,das2018consistent,Petkov}. Simulations provide direct comparison with the EF scratch test.

\section{Results}

\begin{figure}[htb]
\centering
\includegraphics[width=1\columnwidth]{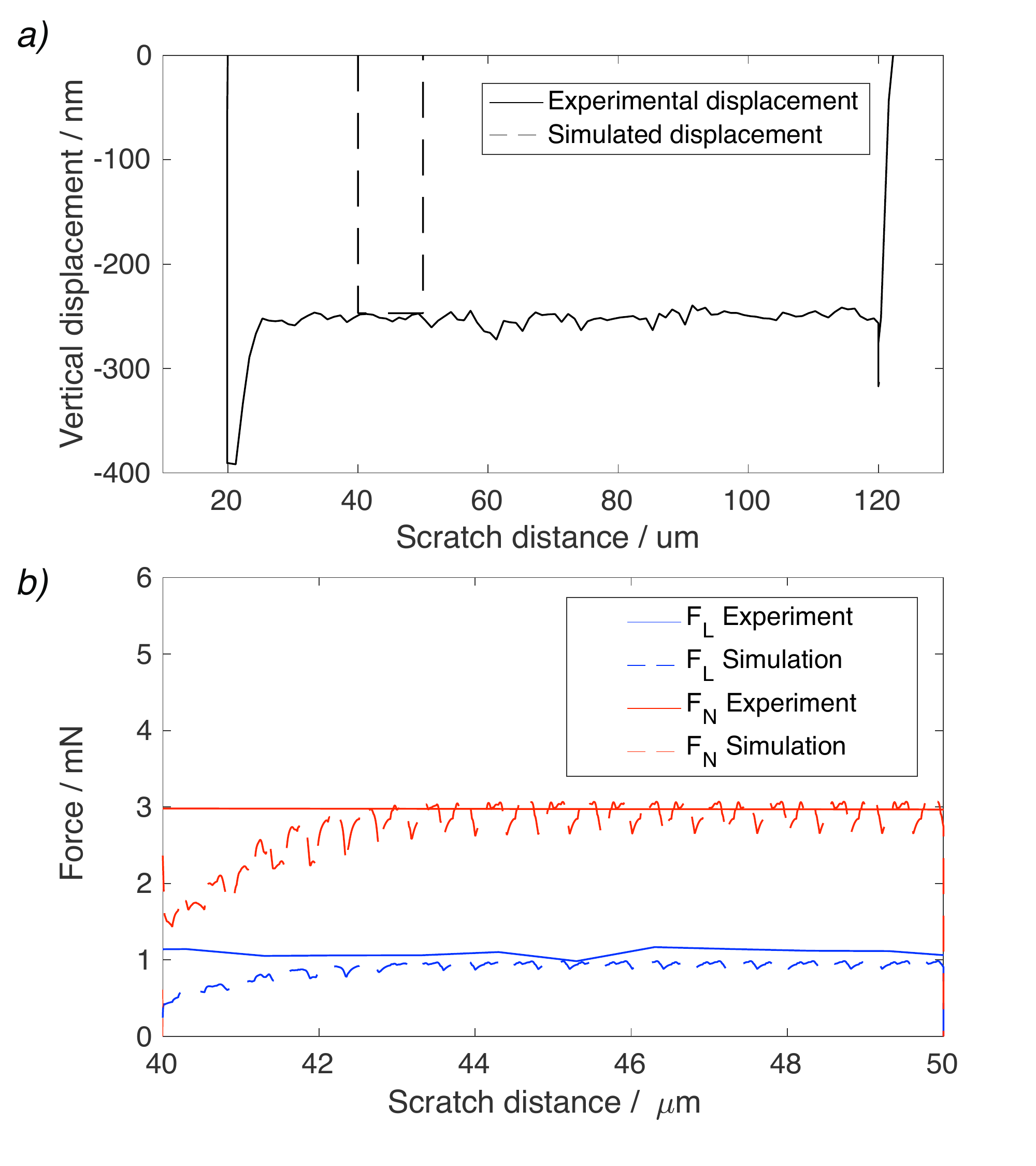}
\caption{a) EF experimental and simulated penetration depth vs. scratch distance. b) corresponding normal force and lateral force vs. scratch distance between $40-50~ \mu$m for experimental and simulated EF scratch.}
\label{Figurex}
\end{figure}

The experimentally measured and simulated vertical displacement, normal force and lateral force are given in Figure \ref{Figurex} for the EF tip orientation. {\color{black} Figure \ref{Figurex}(a) shows the vertical displacement of the tip for the full 100 $\mu$m scratch length measured experimentally. The simulated scratch data is presented for comparison. The vertical dashed lines represent the indenter displacement from zero to the maximum displacement of 247 nm, with no lateral movement, where the tip was loaded and unloaded respectively.} Data from a scratch distance of 40-50 $\mu$m are shown for the forces (Figure \ref{Figurex}(b)), where the experimental scratch had reached a steady state and the normal force was maintained at 3 mN without any influence of loading and unloading of the indenter. As the simulated scratch was only 10 $\mu$m long, this is compared with the steady state region, between 40-50 $\mu$m, of the experimental scratch data. The oscillations that appear in the simulated data are an artefact with a wavelength defined by the the node spacing $w_1$ and the initial 2 $\mu$m can be interpreted as the settling in portion of the scratch. Once steady state scratch deformation is achieved, there is excellent agreement between the experimental and simulated normal and lateral forces.

\begin{figure*}[htb]
\centering
\includegraphics[width=1\linewidth]{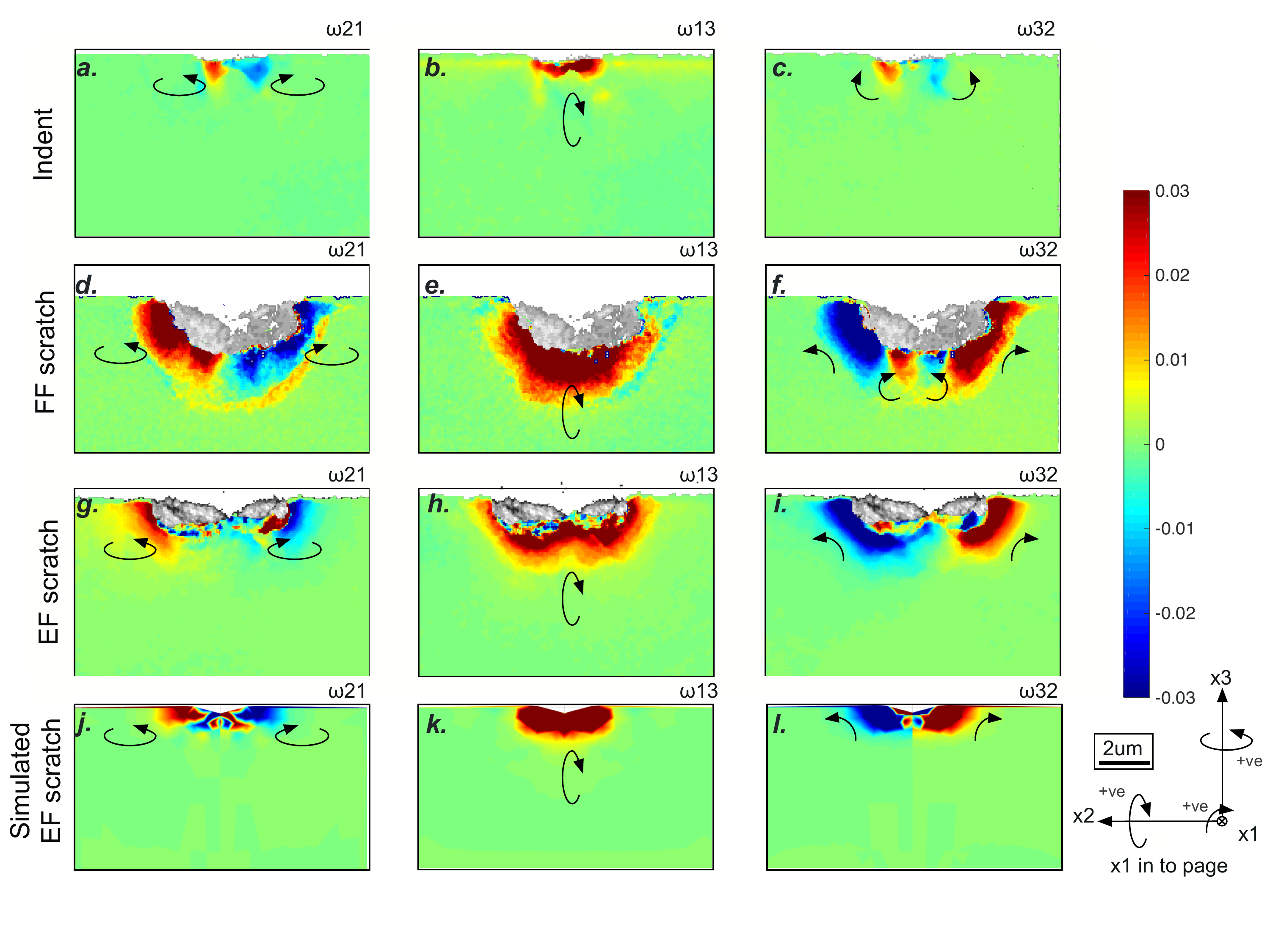}
\caption{Lattice rotation field maps for the indentation section and the scratch sections A (see Figure \ref{Figure1}) about the x$_3$, x$_2$ and x$_1$ axis; $\omega_{21}$, $\omega_{13}$ and $\omega_{32}$ respectively. a, b and c are rotation fields from the indentation, d, e and f are the FF scratch (100) cross section rotation field, g, h and i are the corresponding EF scratch rotation field. j, k and l show the simulated rotation fields of the EF scratch. The colour code shows the lattice rotation in radians. Lattice rotation maps are overlaid on the greyscale image quality maps from EBSD. Scaling is identical for all diagrams.}
\label{fig:x1}
\end{figure*}

\begin{figure*}[htb]
\centering
\includegraphics[width=1\linewidth]{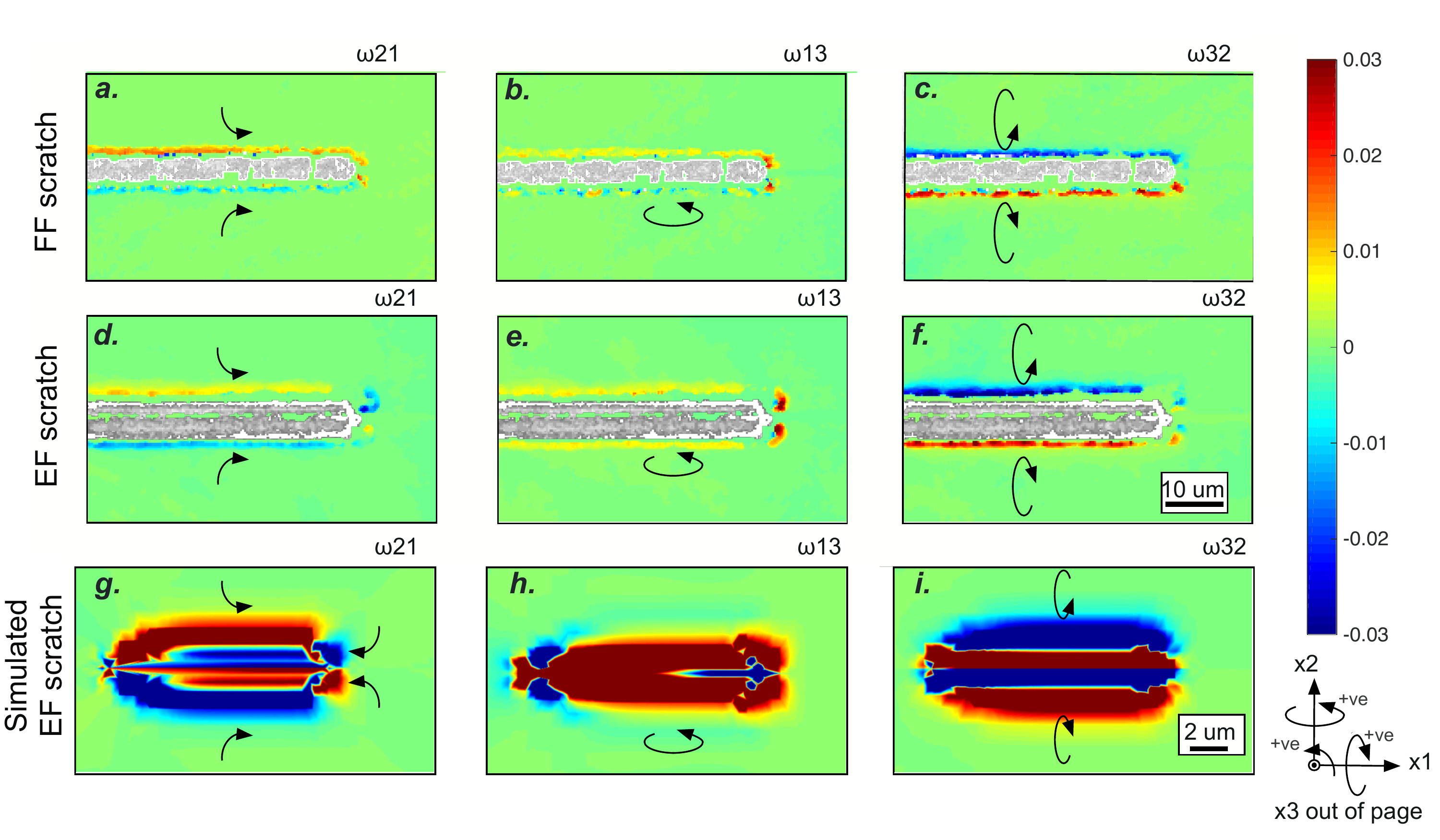}
\caption{Lattice rotation field maps, experimental and simulated, of the free (001) surface about the x$_3$, x$_2$ and x$_1$ axis. Colour code shows the lattice rotation in radians. Lattice rotation maps are overlaid on the greyscale image quality maps from EBSD. Scaling is different between experiment and simulation, refer to scale bars.}
\label{fig:x3}
\end{figure*}

\begin{figure*}[htb]
\centering
\includegraphics[width=1\linewidth]{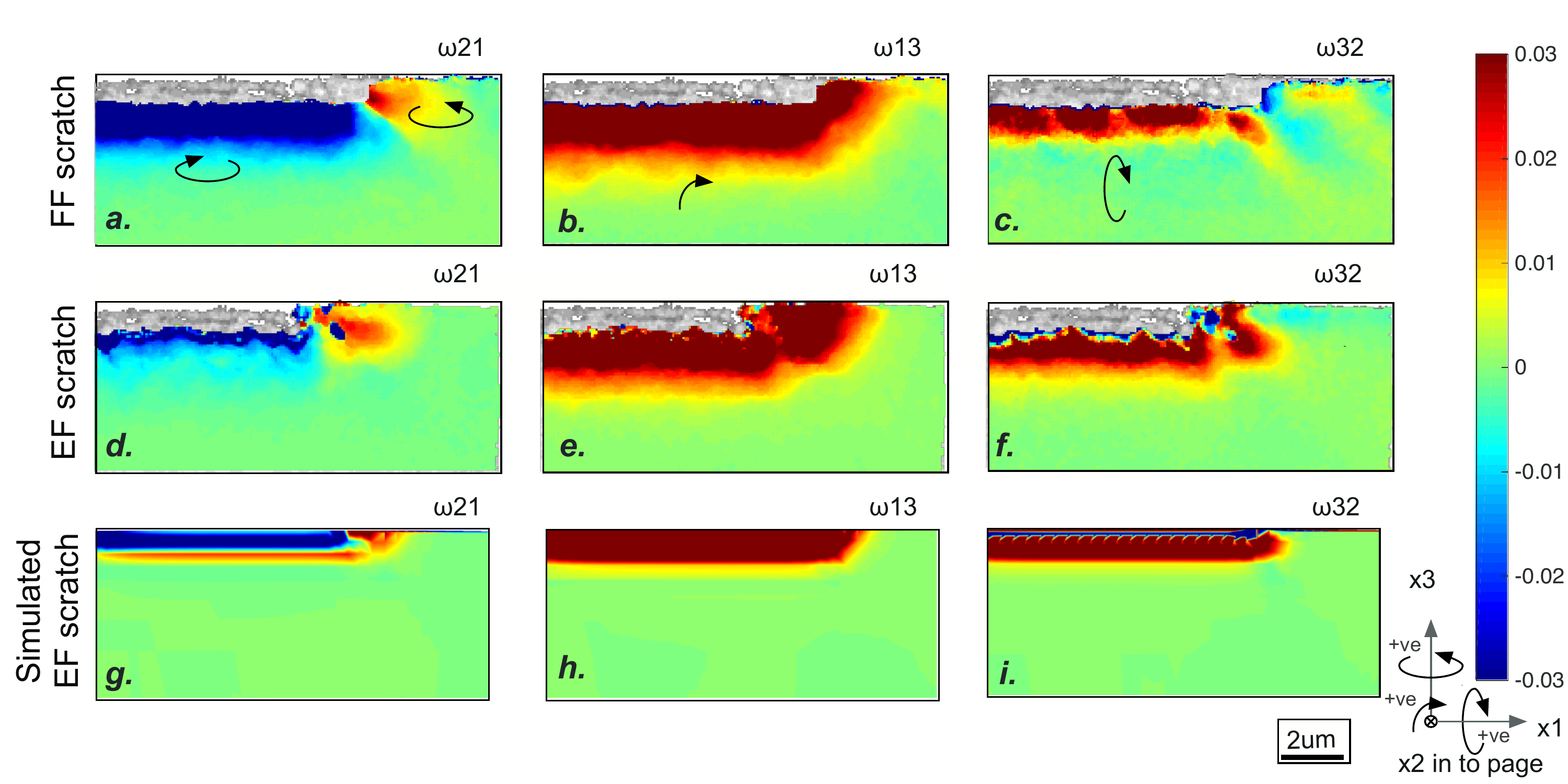}
\caption{Lattice rotation field maps for scratch section B about the x$_3$, x$_2$ and x$_1$ axis. The colour code shows the lattice rotation in radians. Lattice rotation maps are overlaid on the greyscale image quality maps from EBSD. Scaling is identical for all diagrams.}
\label{fig:x2}
\end{figure*}

Experimental and simulated lattice rotation fields, $\omega_{21}$, $\omega_{13}$ and $\omega_{32}$ are shown in Figures \ref{fig:x1}-\ref{fig:x2}. {\color{black}These fields correspond to rotations about the x$_3$, x$_2$ and x$_1$ axis respectively (the crystallographic orientation is shown in Figure \ref{Figure1}(b)).} The colour code represents the lattice rotation in radians; the scale has been confined to a magnitude of 0.03 radians ($1.7^\circ$) to enable a clearer visualisation of the shape and sense of the rotation fields. Owing to a combination of edge effects, highly localised deformation and milling-induced curvature, there was insufficient overlap between the captured EBSD patterns and the reference pattern in regions closest to the indenter. As a result, HR-EBSD was unable to compute the lattice rotation fields in these regions. To indicate the actual surface of the sample, lattice rotation maps are overlaid on the greyscale image quality map from EBSD. {\color{black}Throughout this work, we refer to a positive lattice rotation as an anticlockwise rotation about an axis when looking down the axis towards the origin.}   

Figures \ref{fig:x1}(a)-(c) show the lattice rotation fields for the indentation cross section, measured experimentally using HR-EBSD. Figure \ref{fig:x1}(d)-(f) are the equivalent fields for the FF scratch (100) cross section (indicated A in Figure \ref{Figure1}) and Figures \ref{fig:x1}(g)-(i) are the corresponding fields from the EF scratch. Finally, Figures \ref{fig:x1}(j)-(k) show the corresponding simulated rotation fields for the EF scratch which was taken from a scratch distance of 5 $\mu$m.

From the experimental maps, it is clear that the {\color{black}lattice rotation field} below a scratch extends significantly further than beneath an indentation created under the same normal force and that the FF scratch has a larger deformed region (plastic zone) than the EF scratch. The faceted Berkovich indenter creates a deformation field that takes a distinctive double-lobed form, with a steep maxima in the rotations in regions tangent to the indenter facets - the orientation of these facets with respect to the loading axis is different for each experiment and this is reflected in the shape of the plastic zone. In the indentation and EF oriented scratch, the double-lobe shape is more prominent. {\color{black}In the EF scratch the tip is oriented such that a sharp edge leads the deformation, whereas in the FF oriented scratch a flat facet drives the deformation. For the indentation alone, the very sharp point of the indenter is driving deformation vertically.}

In order to describe the rotation fields, it is assumed that material rotates about the apex of the indenter tip. The indentation field can be interpreted as the lattice rotating toward the central loading axis of the indentation for $\omega_{21}$ and $\omega_{13}$ (Figure \ref{fig:x1}(a) and (b)); $\omega_{32}$ shows the lattice rotates towards the free surface and towards the centre of the indentation (Figure \ref{fig:x1}(c)). Note that for the indentation, $\omega_{21}$ and $\omega_{13}$ (Figure \ref{fig:x1}(a) and (b)) are dependent on the location of the section. Although every effort was made to prepare this section across the centre of the indentation, experimentally this is challenging when using a FIB to prepare cross-sections, and it is likely that it is slightly off centre (by approximately 50 - 150 nm).

{\color{black} The $\omega_{21}$ and $\omega_{13}$ rotation fields in both scratch experiments show that the lattice rotates about the indenter apex, with a positive rotation about the x$_2$ axis perpendicular to the scratch direction.} The in-plane rotation fields $\omega_{32}$ (Figure \ref{fig:x1}(f) and (i)) show that in scratch, the lattice rotates in the opposite direction to that observed for indentation. For the FF case (Figure \ref{fig:x1}(f)) there is an inner region of counter rotation. A region of counter rotation can also be identified in the EF rotation field (Figure \ref{fig:x1}(i)) although it is less pronounced. The model is able to accurately simulate then sign of the experimentally observed deformation-induced rotation field, including the inner zone of counter rotation. Although the simulated plastic zone size is smaller than the experiment, the double-lobed shape is reproduced. The simulation also provides additional information regarding the lattice rotations close to the indenter apex, where experimental data could not be obtained. 

Similar rotation field maps of the scratch surface are given in Figure \ref{fig:x3} about the same axes. On the free surface, the rotation fields correspond with that observed subsurface with a change in rotation sense either side of the scratch track (in $\omega_{21}$ and $\omega_{32}$) and a positive lattice rotation about the axis perpendicular to the direction of travel, [100], in $\omega_{13}$. The experimental scratch width is approximately 4 $\mu$m, which is in accordance with that produced from the simulation. Note that in Figure \ref{fig:x3}, the experimental scratch width appears wider due to the lack of experimental HR-EBSD data close to the scratch edges, due to the poor quality EBSD patterns. At the end of the scratch track there is considerable deformation from piled-up material. For the EF case, the sign of the in-plane rotation $\omega_{21}$ changes at the end of the scratch (Figure \ref{fig:x3}(d) and (g)). This can be more clearly seen in Figure \ref{fig:x2}, which shows the experimental and simulated rotation fields from scratch section B, the (010) plane. As before with indentation, the rotation fields shown for scratch section B are dependent on the precise location of where the cross section was prepared, and although the aim was to target the very centre of the scratch track, experimentally this is challenging and it is likely that the slice was taken slightly off centre. The exact offset for the experimental cross-section is unknown however it is assumed to be within the range of 50 nm-200 nm. Figures \ref{fig:x2}(g)-(i) show the simulated cross-sections offset by 100 nm from the centre of the scratch for comparison. The discrepancy between the experimental and simulated  $\omega_{21}$ rotation field is likely due to this uncertainty.
\begin{figure*}[htb]
\centering
\includegraphics[width=0.9\linewidth]{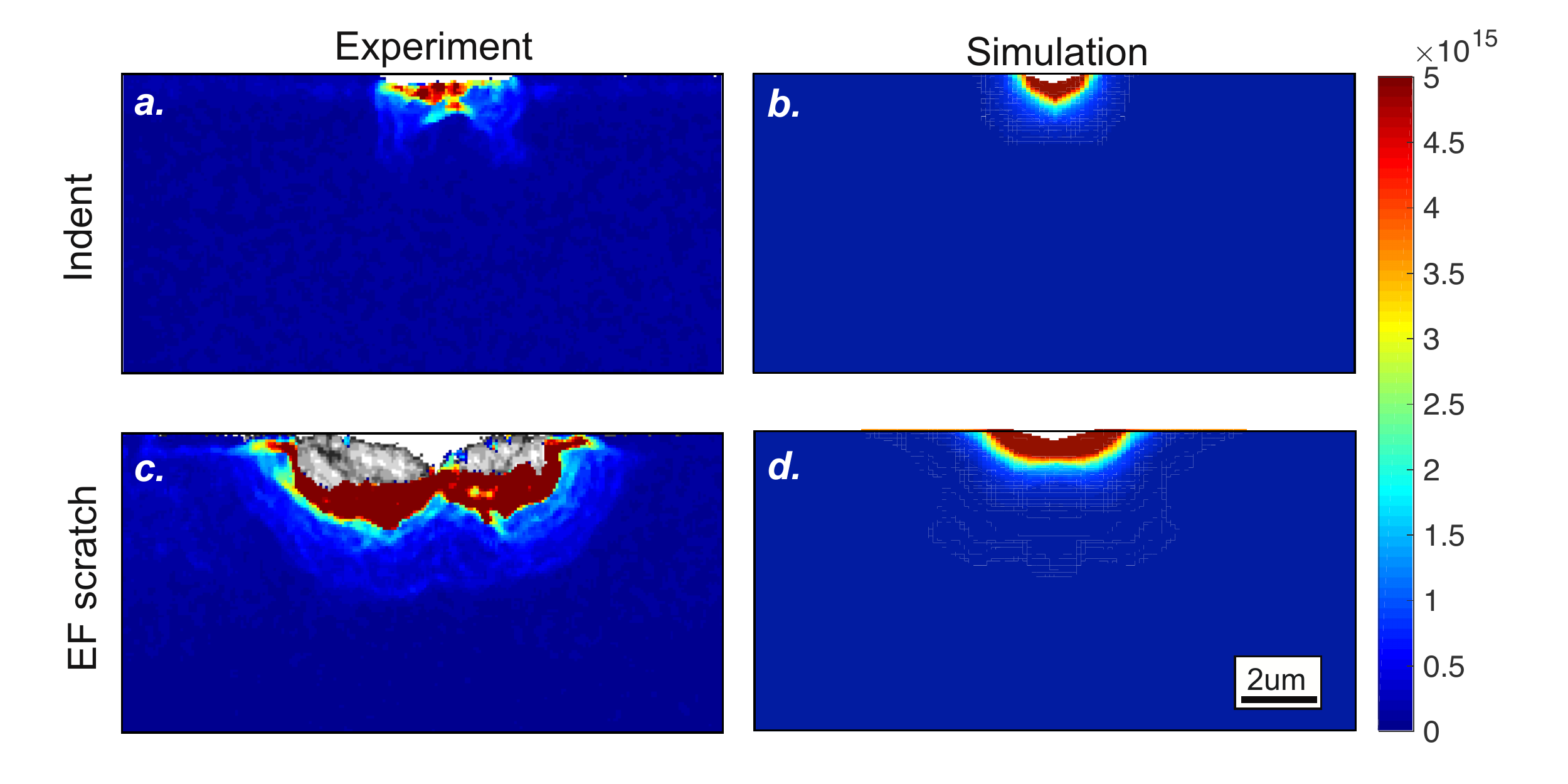}
\caption{GND density map for the indentation (a) and EF scratch section A (b) calculated experimentally using HR-EBSD. Corresponding total dislocation density from the CPFE simulation for the indentation step (b) and the steady state scratch cross-section (d). Colour code shows the dislocation density per $m^{-2}$.}
\label{fig:GND}
\end{figure*}

Figure \ref{fig:GND} compares the experimental GND density field measured using HR-EBSD, with that of the  dislocation density calculated by the model, by numerical integration of $\dot{\rho}$ (defined in equation (\ref{eq:rhossd})) over time. The magnitude and distribution of dislocation density calculated by the model is representative of that measured for the experiment, which is remarkable given the simple form of the dislocation density evolution and hardening laws used in the model. The qualitative agreement between simulated and experimentally measured dislocation density plots supports both the assumption that dislocation multiplication is approximately proportional to effective plastic strain, and the use of the Taylor hardening law in the simulation.

\section{Discussion}
 
This work uses HR-EBSD to study the local deformation field around nanoscale experiments. The multiple views of the scratch rotation field, presented in the experimental results enables a more comprehensive investigation of the volume of deformed material surrounding the scratch. However, it remains limited to a snapshot, postmortem analysis and is not sufficient to measure deformation in the regions where the largest rotations occur, close to the indenter apex. The 3-D rotation fields predicted using a crystal plasticity simulation provide a real-time visualisation of the deformation field during scratch formation and are able to predict the deformation close to the indenter, where experimental data is missing. Hence in this work, the experiments and simulations are complementary to each other, and enable a more complete investigation. The CPFE model is able to accurately predict the rotation field for the EF tip orientation in terms of sense and axis. The accurate simulation of the normal and lateral forces, relies on a correct coefficient of friction which is determined from the experimental data as outlined in Appendix A. The most commonly reported coefficient of friction for scratch tests uses the ratio between the lateral and normal force, however, this has been found to be non-physical when used in the model. A new coefficient of friction, based on the resolved forces acting on the facets of the tip (see Appendix A) is used in the model. This approach is more physically appropriate as it is based on the geometry of the tip and the surfaces in contact and is able to accurately predict the experimental forces. Figure \ref{appendix1} (b) shows the experimentally measured friction coefficients, defined by the conventional method and the new proposed method; incorporating the resolved forces on the facets of the indenter reduces the friction coefficient and produces a coefficient independent of tip orientation. This method can be applied to any pyramidal indenter with a known geometry. Compared to the model, a larger plastic zone is observed in the experimental measurements of rotations and dislocation density fields surrounding the scratch. This difference may be the effect of the indenter rounding in the experiment (20 $nm$ tip radius) which would displace more material than the perfect tip used in the model and/or heterogeneous material properties in the surface layer of the sample as a result of sample preparation induced damage.

Using the results obtained experimentally and by crystal plasticity, the rotation fields can be summarised as follows. For indentation, the lattice rotates about the indenter apex towards the central indenter loading axis. This can be qualitatively understood in terms of the material that must be displaced by the indenter immediately below it, towards the surface to create pile-up around the indentation. The rotation fields $\omega_{21}$ and $\omega_{13}$ measured experimentally, suggest that the centre of the indentation would be further along the x$_1$ axis, into the page of Figure \ref{fig:x1}(a),(b) and (c). 

The sign of the rotation fields around the scratch experiments broadly follow a similar pattern for both the EF and FF tip orientations and can be described by two simultaneous mechanisms. Firstly, the lattice rotates anticlockwise about the x$_2$, as the indenter `pulls' the surrounding lattice along with it in the direction it is traversing. The second mechanism causes the lattice to rotate about the indenter away from the centre of the normal loading axis of the indenter, represented in the $\omega_{32}$ rotation field, which is the opposite sense to that observed in the indentation. This is the most striking difference observed between the indentation and scratch rotation fields in the experimental data.  

The differences in the shape of the plastic zone between the two scratch tip geometries might be explained in terms of the way the material ahead of the indenter is displaced around the tip for each orientation. In the FF tip orientation, the most efficient way would be for material to move underneath the tip, whereas in the EF orientation the angled facets would assist displacement of material laterally around the tip. The exact displacements are more complex and would involve multiple directions for both tip orientations however this simple analogy could describe the differences between the two scratch tip geometries. Additionally, the FF tip orientation has a higher lateral component in the resolved force, compared to the EF tip orientation (see equations in \ref{sec:appendix}) which would result in more work being done on the material, and as a result more plasticity.

\begin{figure*}[htb]
\centering
\includegraphics[width=0.7\linewidth]{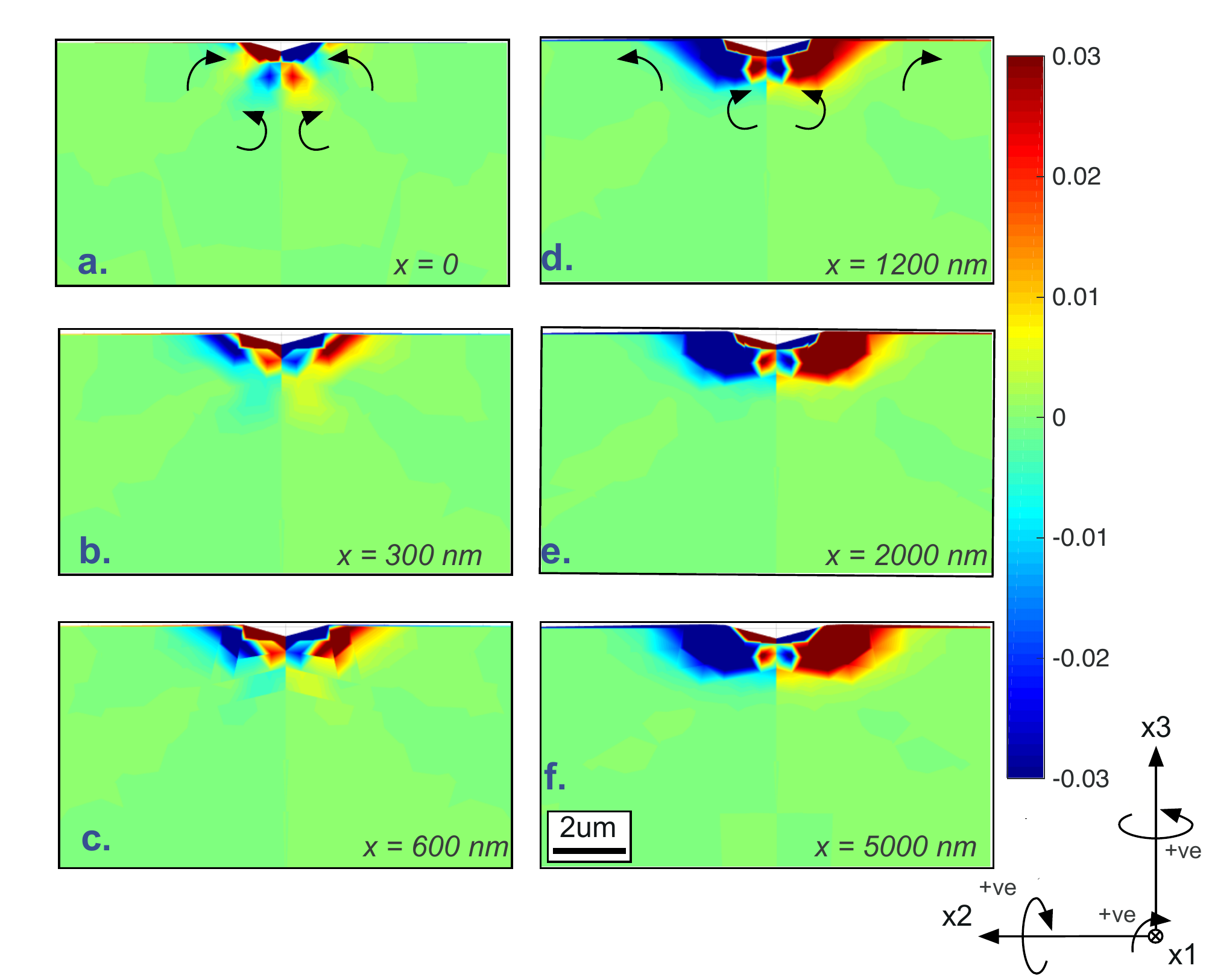}
\caption{Simulated $\omega_{32}$ rotation field, for a set of successive planes through the scratch, parallel to the (100) plane (i.e. parallel to scratch section A). The colour code shows the lattice rotation in radians.}
\label{fig:w12}
\end{figure*}
As the simulation modelled an indentation step followed by the scratch step, analogous to the experiment, it is possible to investigate the mechanics of scratch formation, as the loading state transitions from a static indentation to a steady state scratch. This will aid the interpretation of differences observed between the indentation and scratch experiments. In Figure \ref{fig:w12}, the simulated $\omega_{32}$ rotation field, for a set of successive planes throughout the scratch, parallel to the (100) plane (i.e. parallel to scratch section A) are given. Figure \ref{fig:w12}(a), where the indenter is solely under a normal load (i.e. static indentation), shows a rotation field with four lobes, where the zones close to the surface have the same rotation sense to that observed in the experimental indentation rotation (Figure \ref{fig:x1}(c)). {\color{black}This four-lobed rotation field has been observed for static indentations using a wedge indenter \cite{kysar2007high} and spherical indentations \cite{gao2015lattice, cackett2019spherical}.} As the scratch begins to traverse laterally (Figure \ref{fig:w12}(b)), an additional outer region of counter rotation begins to develop at the surface, whilst the four-lobed indentation rotation zone becomes further confined. As the scratch progresses, (Figure \ref{fig:w12}(c)-(f)) the outer rotation zone expands further and dominates the plastic deformation field. {\color{black} This abrupt change in rotation direction, could correspond with the tribologically induced discontinuity, known as a dislocation trace line, ubiquitously observed in various material systems. The origin of this has been attributed to the reorganisation of dislocations beneath a sliding contact as a result of the in-plane shear component of the stress field \cite{haug2020early, Greiner2016}. It appears that by incorporating a lateral component of force, a strongly inhomegeneous stress state subsurface, different to that under a statically loaded indent, activates slip systems that cause an outer rotation field to be formed whilst constraining the inner rotation field.} Referring back to the scratch displacement profile in Figure \ref{Figure1}(a) it can be seen that in the early stages of the scratch formation, the displacement initially reaches a maximum of approximately $\sim$400 nm, where the maximum normal force of 3 mN is supported by the indentation alone. When the tip begins to traverse laterally, and the outer deformation field begins to form, the 3 mN is only enough force to produce a displacement of $\sim$250 nm. Hence the indenter rises up until it reaches a steady state penetration depth. Although the model uses a displacement controlled scratch step to aid numerical stability, the same mechanisms are observed in Figure \ref{Figurex}. The initial indentation to a target depth of 247 nm requires a normal force of 1.5 mN, as the scratch progresses laterally, and the outer deformation field is formed, the normal force must increase in line with the experiment to maintain the constant penetration depth.

Hence this work highlights that with a reasonable simple variation in the test, i.e. incorporating lateral movement of the tip, the deformation field is significantly different and more plasticity is induced in the material. An outer rotation zone, of opposite sense, forms in addition to the rotation field created by indentation alone. This implies that an indentation process does not have all of the components to describe the deformation associated with a sliding contact. As a result, in terms of predicting the wear resistance of materials, indentation hardness may not be the most suitable measure. Scratch hardness, based on scratch testing will provide a significantly closer representation of the deformation associated with a sliding contact. 

 \section{Conclusions}
 
 We present an investigation of the deformation field in the vicinity of a controlled scratch using a sharp (EF) and relatively blunt (FF) indenter orientation, in a Cu single crystal, for direct comparison with a static indentation. CPFE was calibrated with only 3 parameters and reproduced the normal and lateral forces, as well as all 3 rotation fields on all 3 principal planes. A simple hardening law was found sufficient and gave remarkable agreement with the measured GND density. This combined experimental-modelling approach provides a more complete understanding of the nanoscratch formation. The main conclusions are as follows: 
 
 \begin{itemize}
     \item By applying the same normal force, the three experiments show different lattice rotation fields in terms of the morphology of all three rotations based on the direction of loading with respect to the tip orientation. The biggest difference in the lattice rotation is observed in the sign of the $\omega_{32}$ rotation, which is reversed for indentation and scratch.
     
     \item By simply incorporating lateral movement of the tip, the deformation field is distinctively different to indentation. This implies that indentation, and therefore hardness calculated from indentation alone, cannot fully capture the deformation associated with a sliding contact. Scratch hardness may provide a more appropriate predictor of wear, as it is able to capture all components of deformation associated with a sliding contact.   
     
     \item  The commonly reported friction coefficient, $F_L/F_N$, is not specific to the geometry of the contact and therefore is not sufficient to use in the CPFE model to accurately predict the normal and lateral forces. A more appropriate methodology for calculating the friction coefficient, from nanoscratch, is proposed. 
 \end{itemize}

\section*{Acknowledgements}

AK and ET acknowledges support from the Engineering and Physical Sciences Research Council under Fellowship grants EP/R030537/1 and EP/N007239/1 and Platform grant EP/P001645/1. CH also acknowledges funding by the UKRI Energy Programme (Grant No. EP/T012250/1). We are grateful for use of characterisation facilities within the David Cockayne Centre for Electron Microscopy, Department of Materials which has benefited from financial support provided by the Henry Royce Institute (Grant ref EP/R010145/1)

\section*{Author Contributions }

AK undertook the experimental work (indentation/scratch experiments, FIB and HR-EBSD). ET and CH performed the CPFEM simulations. SVH and AJW advised on the indentation/scratch testing and EBSD respectively. All authors contributed to interpretation of the results. AK led drafting of the manuscript, with all authors contributing and agreeing this final version. 

\appendix

\section{Calculating the friction coefficient from experimental data}
\label{sec:appendix}

The coefficient of friction used in the model is given by Equation (\ref{coeff}) where $F_T$ is the tangential force parallel to the indenter facets and $F_R$ is the resolved force perpendicular to the indenter facets (see Figure \ref{appendix1}). $F_R$ and $F_T$ are given by Equations (\ref{eq:FR}) and (\ref{eq:FT}) respectively for the EF tip geometry. The authors note an error in the equation for $F_R$ presented in the appendix of \cite{Kareer2016paper2} and (\ref{eq:FR}) is the correct expression. 

\begin{align}
\mu &=F_{T}/F_{R}\label{coeff}\\
F_T &= \frac{F_L}{2}\cos\phi - \frac{F_N}{2}\sin\phi
\label{eq:FT}\\
F_R &= F_X\cos\theta + \frac{F_N}{2}\sin\theta\nonumber\\ &=\frac{F_L}{4}\cos\theta + \frac{F_N}{2}\sin\theta\label{eq:FR}
\end{align}
where $F_X=(F_L/2)\cos(60^\circ)=F_L/4$, $\theta = 65.3^\circ$ and $\phi = 12.95^\circ$ for a Berkovich indenter.
\setcounter{figure}{0} 
\begin{figure}[htb]
\centering
\includegraphics[width=0.9\linewidth]{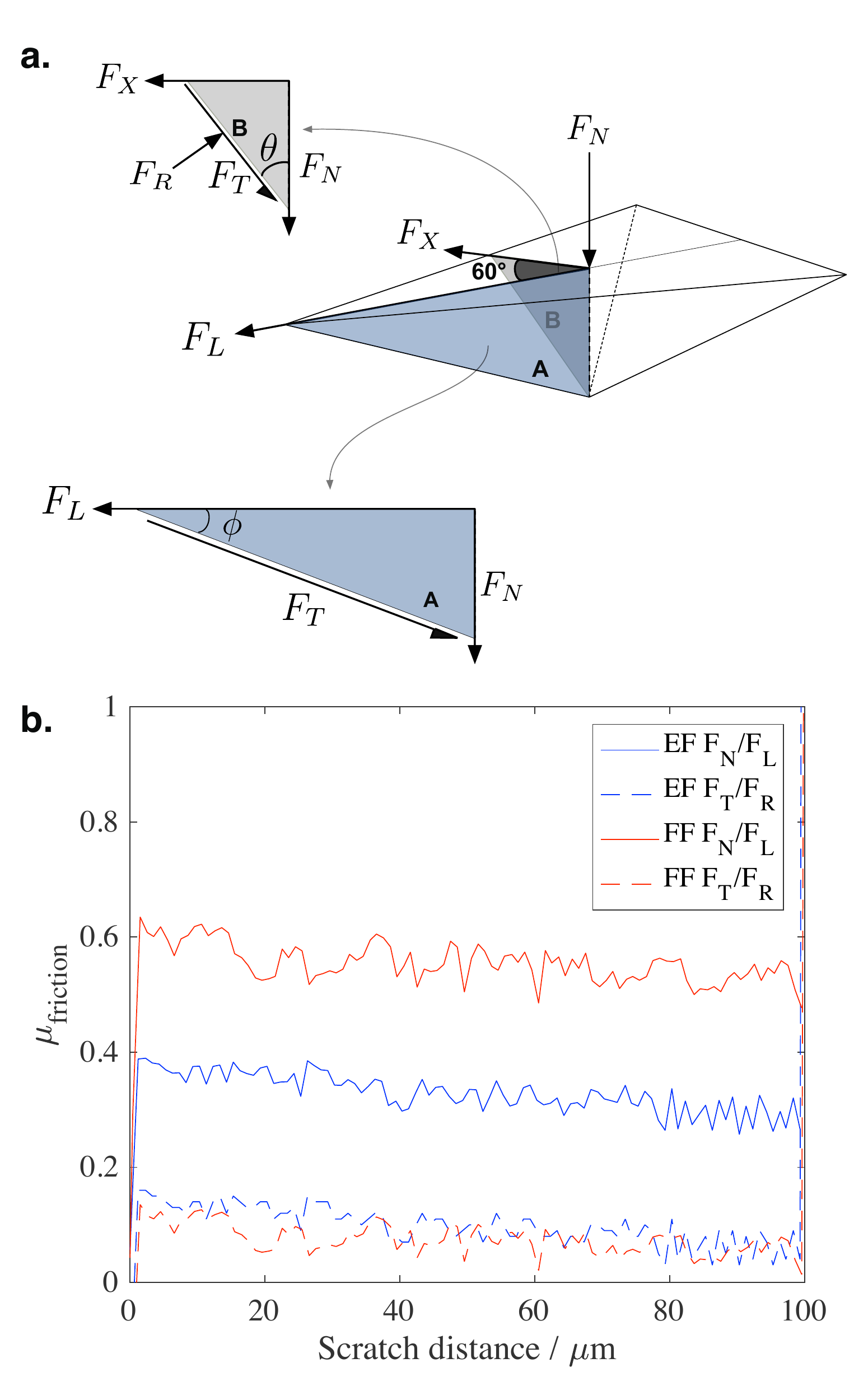}
\caption{(a) Schematic of indenter indicating forces acting on the facets. (b) Friction coefficient as a function of scratch distance using traditional and new definition for both the EF and FF tip orientation}
\label{appendix1}
\end{figure}

In the above equations, it is assumed that the direction of sliding in the EF case is along the scratch direction, however in reality a small component of sliding may be normal to the scratch direction.

In the face forward case (FF) the reaction and tangential forces are
\begin{align}
    F_R = F_L \cos\theta + F_N\sin\theta \\
    F_T = F_L \sin\theta - F_N\cos\theta
\end{align}
and so the friction coefficient (\ref{coeff}) becomes
\begin{equation}
    \mu = \frac{\frac{F_L}{F_N}\tan\theta -1}{\frac{F_L}{F_N}+\tan\theta} = \tan(\theta -\alpha)
\end{equation}
where $\alpha = \tan^{-1}(F_N/F_L)$.

In the FF scratch orientation, the lateral force is along $F_X$ in Figure \ref{appendix1}, with both $F_T$ and $F_R$ on the same plane shown as section B in the figure.


\bibliographystyle{model1-num-names}
\bibliography{scratchpaper.bib}







\end{document}